\pdfoutput=1
\documentclass[prl,twocolumn,showpacs,floatfix]{revtex4-1}
\usepackage{times,amsmath,amssymb,amstext,latexsym,float,graphicx,color}
\usepackage{hyperref}
\hypersetup{colorlinks=true, citecolor=blue, urlcolor=blue, linkcolor=blue}

\begin{document}

\title{Persistent Currents in Toroidal Dipolar Supersolids}

\author{M.~Nilsson Tengstrand}
\email{mikael.nilsson\_tengstrand@matfys.lth.se}
\affiliation{Mathematical Physics and NanoLund, Lund University, Box 118, 22100 Lund, Sweden}

\author{D.~Boholm}
\affiliation{Mathematical Physics and NanoLund, Lund University, Box 118, 22100 Lund, Sweden}

\author{R. Sachdeva}
\affiliation{Mathematical Physics and NanoLund, Lund University, Box 118, 22100 Lund, Sweden}

\author{J. Bengtsson}
\affiliation{Mathematical Physics and NanoLund, Lund University, Box 118, 22100 Lund, Sweden}

\author{S.M.~Reimann}
\affiliation{Mathematical Physics and NanoLund, Lund University, Box 118, 22100 Lund, Sweden}

\date{\today}

\begin{abstract}
We investigate the rotational properties of a dipolar Bose-Einstein condensate trapped in a toroidal geometry. Studying the ground states in the rotating frame and at fixed angular momenta, we observe that the condensate acts in distinctly different ways depending on whether it is in the superfluid or in the supersolid phase. 
We find that intriguingly, the toroidal dipolar condensate  can support a supersolid persistent current which occurs at a local minimum in the ground state energy as a function of angular momentum, where the state has a vortex solution in the superfluid component of the condensate. The decay of this state is prevented by a barrier that in part consists of states where a fraction of the condensate mimics solid-body rotation in a direction \textit{opposite} to that of the vortex. Furthermore, the rotating toroidal supersolid shows hysteretic behavior that is qualitatively different depending on the superfluid fraction of the condensate.
\end{abstract}

\maketitle

In a supersolid quantum many-body state, off-diagonal and diagonal long-range order occur simultaneously~\cite{Leggett1970,Boninsegni2012}. In other words, a solid-like system may have a superfluid fraction of its total mass, leading to interesting new properties that result from the coexistence of solid-body and superfluid behavior. 
Initially, supersolidity was proposed for $^4\hbox{He}$~\cite{Andreev1969,Chester1970,Leggett1970}, 
and  experimental signatures of anomalous rotational properties ~\cite{Kim2004a,Kim2004b} have been vividly debated; for reviews see for example Refs.~\cite{Balibar2010,Boninsegni2012,Chan2013}.
Ultra-cold atomic Bose-Einstein condensates (BECs) are alternative candidates to realize supersolidity;  examples are  BECs with soft-core two-body potentials~\cite{Pomeau1994,Henkel2010,Cinti2010,Saccani2011}, or condensates coupled to two optical cavities~\cite{Leonard2017a,Leonard2017b}. Likewise,
spin-orbit coupled BECs~\cite{Lin2011, Li2016}  can be in a modulated density stripe phase showing supersolid properties~\cite{Li2017}.
%NOW ABOUT DIPOLAR GASES AND DROPLETS
Another prominent example are quantum gases with long-range dipolar interactions~\cite{Lahaye2009,Baranov2012}. Classical ferrofluids are known for their rich pattern formation; when  a paramagnetic fluid is brought into a magnetic field, it self-organizes into surface structures or small droplets in a Rosensweig transition~\cite{Cowley1967}.
Experiments with ultra-cold dipolar gases found similar quantum-ferrofluid patterns deeply in the quantum realm: In dysprosium condensates, a transition from a superfluid to a metastable state was observed where mutually repulsive long-lived droplets form~\cite{Kadau2016,Schmitt2016,FerrierBarbut2016,Wenzel2017}, each consisting of a large number of atoms. A similar stabilization of droplets was also found with erbium~\cite{Chomaz2016}. 
Inducing droplets and subsequently melting them back into a BEC by Feshbach-controlling the short-ranged part of the dipolar atom-atom interactions, a bistability in the gas-droplet transition was found~\cite{Kadau2016}. As dipoles repel each other when side-by-side but become attractive when head-to-tail, droplets form with a spheroidal shape, elongated along the dipolar polarization direction~\cite{FerrierBarbut2016}. 
When held together by an overall radial confinement and brought to overlap interference patterns were found, signaling the droplets' phase coherence~\cite{Schmitt2016,FerrierBarbut2016}.  
The physical origin of the self-binding mechanism for dipolar droplets 
~\cite{Wachtler2016a,Wachtler2016b,Baillie2016,Bisset2016,Baillie2018}
is similar to that in binary Bose gases~\cite{Bulgac2002,Petrov2015,Petrov2016}  realized with potassium in two hyperfine states~\cite{Semeghini2018,Cabrera2018}.  In either case, the stability of the droplets and their critical size is controlled by the balance between a residual mean-field interaction and quantum fluctuation contributions to the total energy, often referred to as the Lee-Huang-Yang corrections~\cite{lhy1957,Lima2011}. 
%THEN SUPERSOLIDS IN DIPOLAR DROPLETS
For the dipolar BEC, phase transitions from a superfluid  to a density-modulated supersolid  phase and to a droplet crystal phase have recently been observed~\cite{Tanzi2019a,Bottcher2019,Chomaz2019,Tanzi2019b}. 
Increasing the ratio between the dipolar and short-range interaction strengths beyond a critical value, the dispersion relation shows the softening of the roton spectrum at a finite momentum which is a precursor of the modulation of the superfluid density. With further increase of this ratio of dipolar and short-range interaction strengths, there is another phase transition from the density modulated supersolid phase to the droplet crystal phase, where the individual droplets possess superfluid character but have no overlap of superfluidity between the neighboring droplets. Higgs amplitude and Nambu-Goldstone modes~\cite{Hertkorn2019,Tanzi2019b,Guo2019,Natale2019} depicting the spontaneous symmetry breaking in the supersolid phase provided further evidence of supersolid behavior. 
One of the most distinct features of superfluidity, however,  is the quantization of circulation and formation of vortices in response to rotation. The  non-classical rotational moment of inertia provides means to quantitatively differentiate the supersolid from the superfluid phase~\cite{Leggett1998}.  The interplay between 
droplet formation and vortices in pancake-shaped traps that were set rotating, was analyzed in Ref.~\cite{Roccuzzo2020}. In a similar setting, the gas expansion upon switching off the trap confinement was modeled in an extended mean-field approach in Ref.~\cite{Ancilotto2020}, suggesting characteristic interference patterns to signal the supersolid phase in contrast to the ballistic expansion of localized droplet states.

Here, we investigate the existence of a persistent current in a dipolar BEC of dysprosium that is confined in a toroidal geometry. 
Such non-trivial trap shapes have been extensively studied theoretically for condensates with contact interactions (see for example Refs. ~\cite{Karkkainen2007,Bargi2010,Cominotti2014,Mateo2015,Roussou2015}) 
or for dipolar gases in the purely superfluid limit~\cite{Abad2010,Abad2011,Malet2011,Adhikari2012,Karabulut2013,Zhang2015}.
Experiments measuring phase slips and persistent currents with $^{23}\mathrm{Na}$ in ring potentials were performed~\cite{Wright2013,Eckel2014,Eckel2018}, 
emphasizing ``atomtronic" applications. 
The existence of a persistent current state in superfluid rings is well known: At $L/N\hbar =1$ (where $L$ is the angular momentum and $N$ the norm)  the energy as a function of  angular momentum has a characteristic sharp v-shaped minimum with negative curvature, protecting the state from its dissipative decay to a non-rotating state by an energy barrier. 
Here, we show that when a dipolar gas of dysprosium in a torus confinement is set rotating,  it acts in a distinctly different way depending on whether it is in the  superfluid or in the supersolid droplet phase, yet exhibiting surprisingly universal behavior.  We find that intriguingly, also as a supersolid, the system does support superflow.  Deep in the superfluid regime, a persistent current occurs in the usual way, with a characteristic local minimum in the ground state energy at $L/N\hbar = 1$.  In the supersolid, likewise a pronounced local minimum is found, however 
now occurring at  $L/N\hbar < 1$ and with positive curvature. 
In this case, the state carries a vortex in the superfluid component of the condensate. 
The decay of this state  is prevented by a barrier that in part consists of states where a component of the BEC mimics solid-body rotation in a direction \textit{opposite} to that of the vortex. Furthermore, when the supersolid is set rotating, it shows hysteretic behavior that is qualitatively different for different values of the superfluid fraction  of the condensate. 
We found that a toroidal confinement is a particularly favorable setup, since the confinement inhibits the lateral repulsion between the droplets, otherwise diminishing the inter-droplet tunneling~\cite{Wenzel2017}.  

\begin{figure}
  \centering
  \includegraphics[width = 0.4\textwidth]{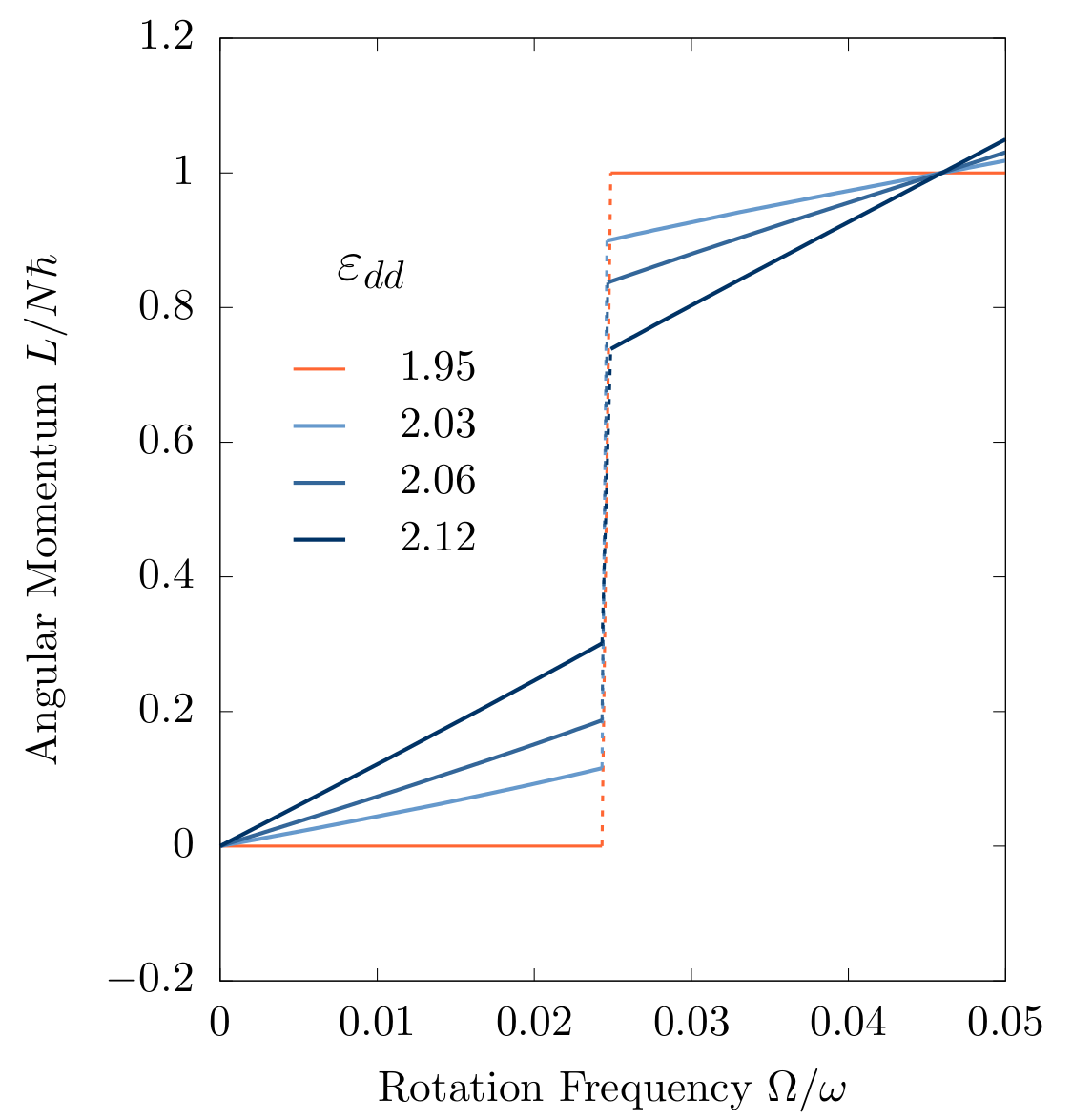}
  \caption{Angular momentum $L/N\hbar $ of the ground state in the rotating frame as a function of rotation frequency for  values of $\varepsilon_\mathrm{dd}$ as indicated by the legend. In the superfluid phase for $\varepsilon _\mathrm{dd}=1.95$ one can observe the typical jump from $L/N\hbar =0$ to $1$ (as here marked by the dashed line). In the supersolid case the linear gain in angular momentum reflects the solid-body response to rotation.}
  \label{fig:lom}
\end{figure}

The dipolar condensate at zero temperature is modeled by a non-local extended Gross-Pitaevskii equation
\begin{multline}\label{gpe}
i\hbar\frac{\partial}{\partial t}\psi(\mathbf{r}, t) = \bigg[-\frac{\hbar^2 \nabla^2}{2m} + V_\mathrm{trap}(\mathbf{r}) + g|\psi(\mathbf{r},t)|^2 + \\ 
\int d\mathbf{r}' V_\mathrm{dd}(\mathbf{r} - \mathbf{r}')|\psi(\mathbf{r}', t)|^2 + \gamma|\psi(\mathbf{r}, t)|^3 \bigg] \psi(\mathbf{r}, t),
\end{multline}
\noindent where $\psi$ is the order parameter normalized to the number of particles $N$, with trapping potential in cylindrical coordinates $V_\mathrm{trap}(\mathbf{r}) = m\omega^2/2[(\rho-\rho_0)^2 + \lambda^2 z^2]$, where $\rho_0$ is the radius of the ring, $\omega$ the axial trapping frequency and $\lambda$ the ratio of transversal and axial trapping frequencies. The contact interaction coupling constant is $g = 4\pi\hbar^2a/m$, where $a$ is the $s$-wave scattering length. The ratio between the dipole and contact interaction strengths is quantified by the dimensionless parameter $\varepsilon_\mathrm{dd} = a_\mathrm{dd}/a$, where $a_\mathrm{dd}=m\mu_0\mu^2/12\pi\hbar^2$ is the dipolar length, $\mu$ is the magnetic moment and $\mu_0$ the permeability in vacuum. The dipolar potential is $V_\mathrm{dd}(\mathbf{r}) = \frac{\mu_0\mu^2}{4\pi}\frac{(1 - 3\cos^2\theta)}{|\mathbf{r}|^3}$, where $\theta$ is the angle between $\mathbf{r}$ and the alignment direction of the dipoles, here taken to be the $z$-direction. The last term is the quantum fluctuation correction term~\cite{Lima2011,Bisset2016}, with $\gamma = \frac{128\sqrt{\pi}}{3}\frac{\hbar^2 a^{5/2}}{m}(1+\frac{3}{2}\varepsilon_\mathrm{dd}^2)$. The ground state is found by solving Eq.~(\ref{gpe}) in imaginary time using the split-step Fourier method. In order to study the system in a rotating frame the term $-\Omega L_z \psi$ (where $\Omega$ is the rotation frequency and $L_z = xp_y - yp_x$) is added to the right-hand side of Eq.~(\ref{gpe}), which is then solved in a similar manner. To obtain the ground state at a fixed angular momentum $L_0$, we instead minimize the quantity $\Tilde{E} = E + C\omega(L - L_0)^2$, where $E$ is the energy corresponding to Eq.~(\ref{gpe}), $L = \int d\mathbf{r}\psi^\ast L_z \psi$ and $C$ a dimensionless positive number that, when large enough, causes the energetic minimum to occur at $L \approx L_0$. (We note here that the numerical solution of Eq.~\ref{gpe} is involved due to many close-lying local minima in the energy surface, which require an extensive sampling over a large number of different initial conditions in order to find the most probable lowest-energy solutions). 
\begin{figure}[H]
\centering
\includegraphics[width = 0.4\textwidth]{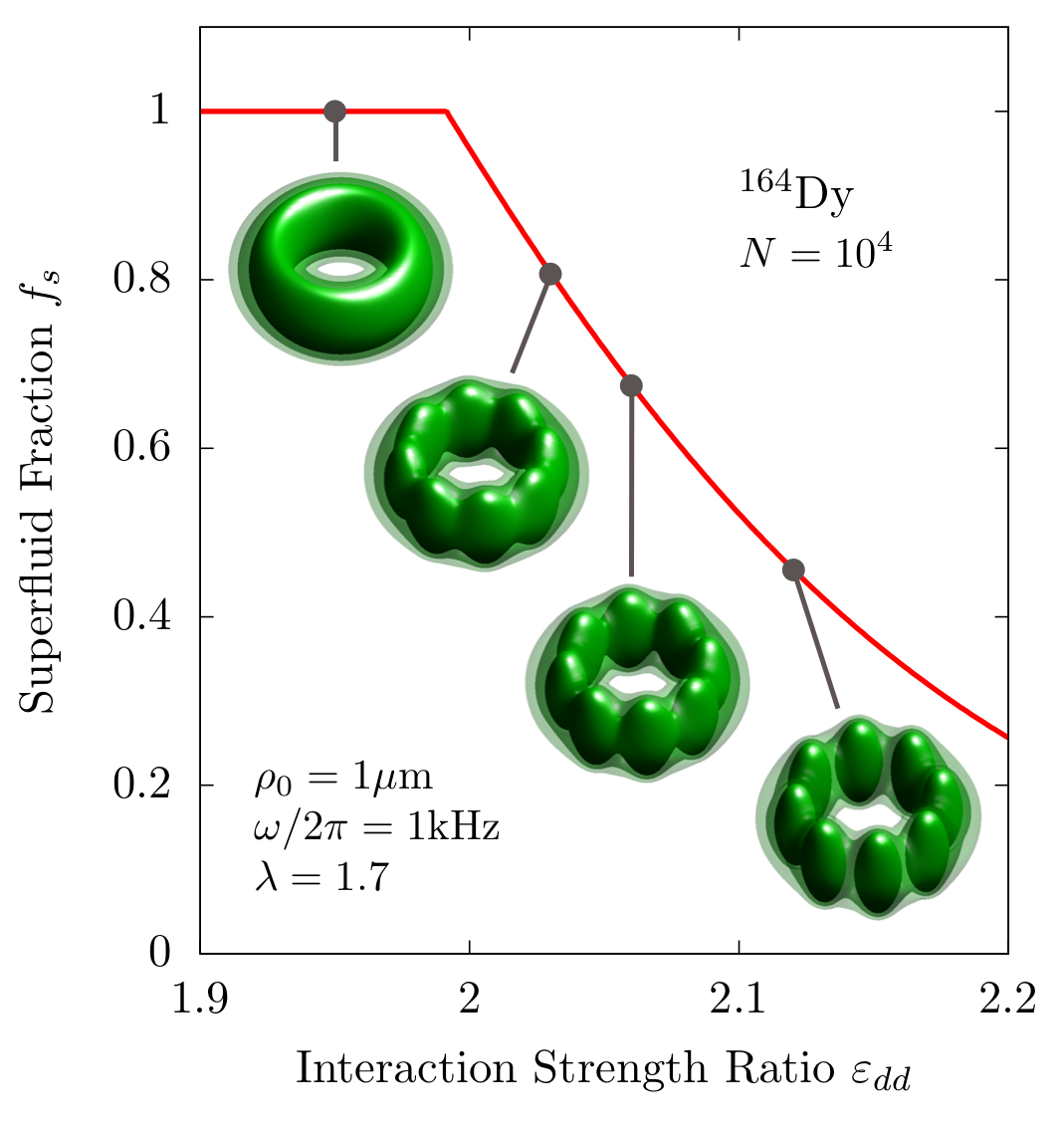}
\caption{Superfluid fraction $f_s$ of a dipolar gas of $^{164}\hbox{Dy}$ atoms 
in a ring-shaped potential of radius $\rho _0= 1~\mathrm{\mu m}$, $\omega /2\pi =~\mathrm{1kHz} $ and transversal trapping ratio $\lambda = 1.7$,  as a function of the interaction strengths ratio $\varepsilon_\mathrm{dd}$ for a norm of $N=10^4$. 
The insets, from left to right, show density isosurfaces for $\varepsilon_\mathrm{dd} = 1.95, 2.03, 2.06$ and  $2.12$, taken at density values $2\times 10^{-4}, 0.5\times 10^{-4}$ and $1.0\times 10^{-5}$(in units of $a_{\mathrm{dd}}^{-3}$). The density profile clearly shows the development from the pure superfluid phase (upper left) with superfluid fraction $f_s=1$ to a density modulated supersolid phase with $f_s< 1$ for increasing $\varepsilon _{dd}$.}
\label{fig:sff}
\end{figure}
We exemplify our findings here for a system of $^{164}$Dy atoms with $N = 10^4$ in a trap where $\rho_0 = 1\text{ }\mu\mathrm{m}$, $\omega/2\pi = 1$~kHz and $\lambda = 1.7$. Different phases of the BEC are identified by the superfluid fraction, defined through its non-classical rotational inertia according to $f_s = 1 - I/I_\mathrm{cl}$ ~\cite{Leggett1970,Leggett1998}, where $I_\mathrm{cl} = m\langle \rho^2 \rangle$ is the classical moment of inertia and $I = \lim_{\Omega\to 0}L/\Omega$. 
The numerically calculated superfluid fraction as a function of $\varepsilon_\mathrm{dd}$ is shown in Fig.~\ref{fig:sff}, where the insets show some typical density isosurfaces. It is seen that for some critical value of $\varepsilon_\mathrm{dd}$ the superfluid fraction $f_s$ drops from unity in the pure superfluid case to smaller values $f_s<1$. 
While the density isosurfaces in the purely superfluid case are azimuthally symmetric (see the top inset in Fig.~\ref{fig:sff} for $\varepsilon _\mathrm{dd}=1.95$), the drop in $f_s$ is accompanied by an onset of localization (middle insets for $\varepsilon _\mathrm{dd}=2.03$ and $2.06$), where the dipolar gas in the toroidal confinement begins to form a modulated density distribution as a precursor to droplet formation, in a way similar to what was previously found in a tube with periodic boundary conditions~\cite{Roccuzzo2019}. As the superfluid fraction decreases the density modulation becomes more pronounced, and leads to the characteristic form of spheroidal droplet-like formations~\cite{FerrierBarbut2016,Wenzel2017,Bottcher2019} elongated along the polarization axis of the dipolar gas and submerged into a finite background of superfluid density, indicating the co-existence of off-diagonal and diagonal long-range order. These droplets are clearly visible in the isosurfaces   (see bottom right inset for $\varepsilon _{\mathrm{dd}}=2.12$).  
\begin{figure}
  \centering
  \includegraphics[width = 0.48\textwidth]{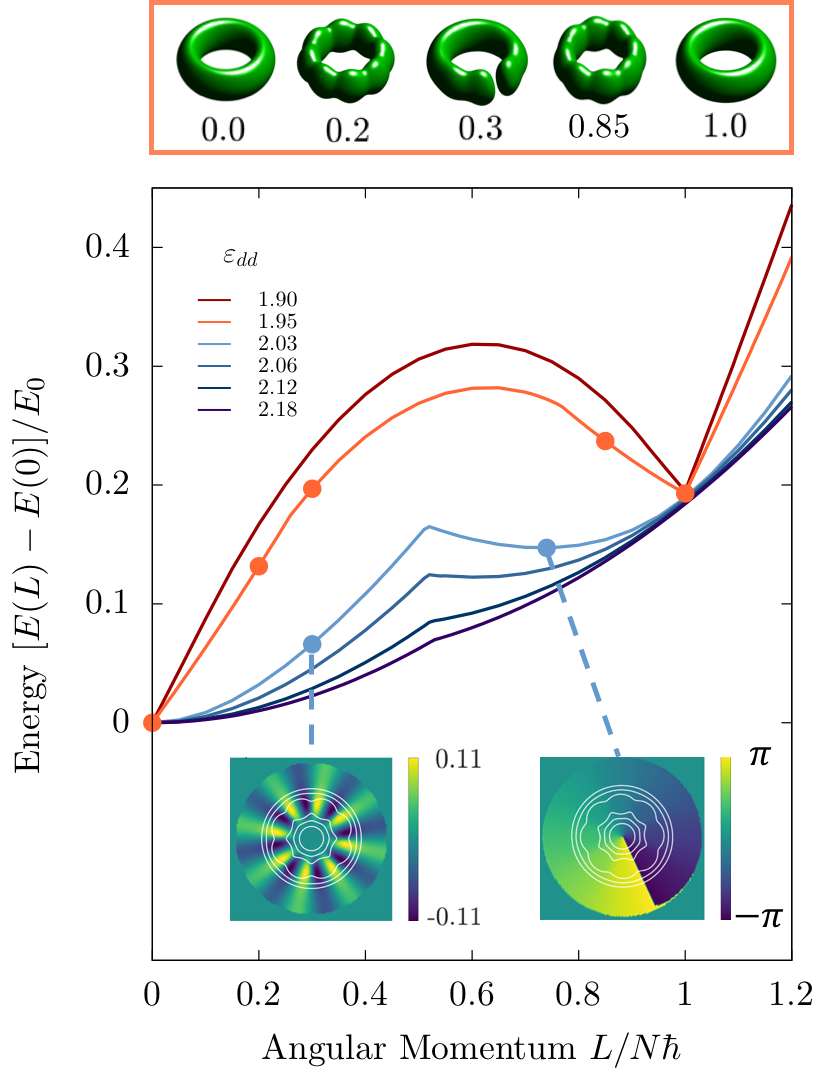}
  \caption{{\it Lower panel:} Ground state energy relative to the non-rotating ground state, $[E(L)-E(0)]/E_0$, where $E_0 = \hbar^2/m a_\mathrm{dd}^2$,
as a function of angular momentum for values of $\varepsilon_\mathrm{dd}$ as given in the legend. In the regime where the superfluid fraction $f_s=1$ (as here shown for the dark red and orange lines), the energy  $E(L)$ develops v-shaped minima when a unit vortex localized at the center of the torus for $L/N\hbar =1$, enabling the system to carry a persistent current.  
In the supersolid regime with $f_s<1$ (blue lines), the minimum shifts to smaller values of $L/N\hbar $, supporting a persistent current in the 
superfluid part. Upon increasing the droplet localization this minimum gradually disappears, and a persistent current can no longer be supported. 
The insets show the phase plots in the ring plane (i.e. at $z=0$), with superimposed density contours (white lines), for $\varepsilon_{dd}$ at $L/N\hbar = 0.3$ and $0.74$.
{\it Upper panel:} Density iso-surfaces for $\varepsilon _\mathrm{dd}=1.95$ (similar to the insets in Fig.~\ref{fig:sff} but taken at a single value of the density  $2\times 10^{-4}a_{\mathrm{dd}}^{-3}$). The sequence shows how in the superfluid phase close to the supersolid transition, the vortex passes through the toroidal condensate, accompanied by a density modulation, until it localizes at the center of the torus 
for angular momentum $L/N\hbar =1$, where azimuthal symmetry is re-established.}
\label{fig:el}
\end{figure}
Let us now determine the ground state energy as a function of  angular momentum, $E(L)$,  for different values of $\varepsilon_\mathrm{dd}$, see Fig.~\ref{fig:el}. 
Deep in the superfluid phase the energy has the  typical negative curvature with the v-shaped local minimum at $L/N\hbar = 1$.
The dark-red line in Fig.~\ref{fig:el} shows $E(L)$ for $\varepsilon _{dd}=1.9$.
Upon increasing $L/N\hbar $ from the non-rotating state, a vortex enters the torus 
from the outside as usual for a toroidal BEC, and becomes localized at the torus center when $L/N\hbar =1$, accompanied by the usual phase jump of $2\pi$, 
similarly to the well-known case of purely superfluid condensates~\cite{Roussou2015}.
Closer to the critical value of $\varepsilon_\mathrm{dd}$, when 
approaching the supersolid regime,  as here shown for $\varepsilon_\mathrm{dd} = 1.95$ (orange line), the entry of the vortex becomes qualitatively different. 
We can see from the density isosurfaces shown in the upper panel that for angular momenta either close to the non-rotating state, or the state with one vortex, the ground state shows a density modulation, as visible in the plots for $L/N\hbar = 0.2$ and $L/N\hbar = 0.85$. For these values it is energetically more favorable to add energy in terms of solid-body rotation instead of the usual vortex entry from the outside of the torus. (We note that this vortex entry is accompanied by density modulations, see the isosurfaces for $L/N\hbar =0.3$).
It should be emphasized that in the vicinity of the phase boundary there are many close-lying solutions such that it is difficult to determine the state that is lowest in energy. For increased $\varepsilon_\mathrm{dd}$, 
the condensate enters the supersolid phase, as we have seen in Fig.~\ref{fig:sff} above.  Curiously, when the system becomes supersolid, a distinct local minimum in $E(L)$ for non-zero angular momentum remains, but now at a value $L/N\hbar < 1$ and with a positive curvature. As in the superfluid case, however, this minimum is protected by an energy barrier against dissipative decay to the non-rotating state.   
The insets in the lower panel of Fig.~\ref{fig:el} show the phase and density contours on either side of the energy barrier, at $L/N\hbar =0.3$ and at the position of the minimum at $L/N\hbar =0.74$, respectively. While for lower angular momentum the phase shifts indicate solid-body rotation, at the minimum beyond the energy barrier the system has developed a singly-quantized vortex. The typical phase jump of $2\pi $ occurs simultaneously with the dipolar density modulation. The density contours (white lines) well reflect the density iso-surfaces for the corresponding values of $\varepsilon _\mathrm{dd}$ shown in Fig.~\ref{fig:sff} above.
By increasing $\varepsilon_\mathrm{dd}$ further towards the limit of droplet localization, the local energy minimum with positive curvature eventually disappears, and the energy instead becomes a monotonically increasing function, no longer capable of supporting a persistent current. (The situation in this case becomes similar to a binary droplet with center-of-mass excitation~\cite{Kavoulakis2020}). 
Complementary to the data shown in 
Fig.~\ref{fig:el}, in Fig.~\ref{fig:lom} we also present the angular momentum of the ground state in the rotating frame as a function of the rotation frequency $\Omega $.  In the superfluid regime, the angular momentum displays its familiar step function behavior with the characteristic jump from $L/N\hbar =0$ to $1$ at a critical frequency $\Omega _{\mathrm{crit}}$ upon localization of the vortex at the torus center.
When the system is in the supersolid regime, however, it also shows a linear gain in angular momentum with increased $\Omega $, reflecting its solid-body response to the rotation. 

As a first attempt to understand the results in the supersolid phase we model the condensate as $f_s N$ particles acting as a superfluid and $(1-f_s)N$ particles acting as a classical solid. Considering the energy as a function of angular momentum, the two lowest energetic branches must then for low angular momenta correspond to classical rotation in the solid part, plus either a single, or no vortex in the superfluid. For the branch without a vortex the energy relevant to the angular momentum considerations here is thus $E_1 = L^2/2I_\mathrm{solid}$, where $I_\mathrm{solid}=(1-f_s)Nm\rho_0^2$ (using $\langle \rho^2\rangle \approx \rho_0^2$) is the approximate moment of inertia of the solid component. Similarly, the energy for the single-vortex branch is $E_2 = (L-f_sN\hbar)^2/2I_\mathrm{solid} + E_V$, where $E_V$ is the energy difference between the states with zero and one vortex for a superfluid with $f_s N$ particles. This energy can be estimated by considering the superfluid component alone, making the ansatz
\begin{equation*}
\psi_\mathrm{sf}(\mathbf{r}) = \sqrt{\frac{f_s N m\omega\sqrt{\lambda}}{2\pi^2\rho_0\hbar}}e^{-m\omega(\rho-\rho_0)^2/2\hbar}e^{-m\lambda\omega z^2/2\hbar}e^{i\ell_\mathrm{sf}\varphi},
\end{equation*}
\noindent where $\ell_\mathrm{sf}$ is the angular momentum per particle in the superfluid. It is further assumed that $\psi_\mathrm{sf}$ goes to zero at the origin when $\ell_\mathrm{sf}\neq0$ and that $\rho_0 \gg \sqrt{\hbar/m\omega}$. The energy is then found to be $E_V = f_sN\hbar^2 / 2m\rho_0^2$, which implies that the two energetic branches intersect at $L/N\hbar = 1/2$, i.e the ground state in the supersolid phase always changes between zero and one vortex at this value. This in turn means that $E(L)$ has a local minimum at $L/N\hbar = f_s$ whenever $f_s > 1/2$, and there can thus exist a persistent current also when the BEC is in the supersolid phase, which agrees well with the numerical data shown in Fig.~\ref{fig:el}. With this interpretation the energetic barrier that prevents the decay of the vortex state consists of two parts: One with just solid-body rotation (to the left of the local maximum), and the other with a single vortex plus solid-body rotation in the {\it reverse} direction (to the right of the local maximum). These peculiar counter-rotating states are, however, not rotational ground states of the system, as adding the energy term $-\Omega L$  never puts them lower in energy than the single-vortex state with no solid-body rotation. The critical rotation frequency for the first vortex (in both the superfluid and supersolid phase) can within our model be estimated to be $\Omega_\mathrm{crit} = \hbar/2m\rho_0^2$.  
Remarkably, this critical frequency is independent of the superfluid fraction, in agreement with the data shown in Fig.~\ref{fig:lom}. The value predicted by our model $\Omega_\mathrm{crit}/\omega \approx 0.03$ overestimates the critical frequency compared with the numerical results due to a deviation from the approximation $\langle\rho^2\rangle  \approx \rho_0^2$ for our particular parameters.
\begin{figure}
  \centering
  \includegraphics[width = 0.5\textwidth]{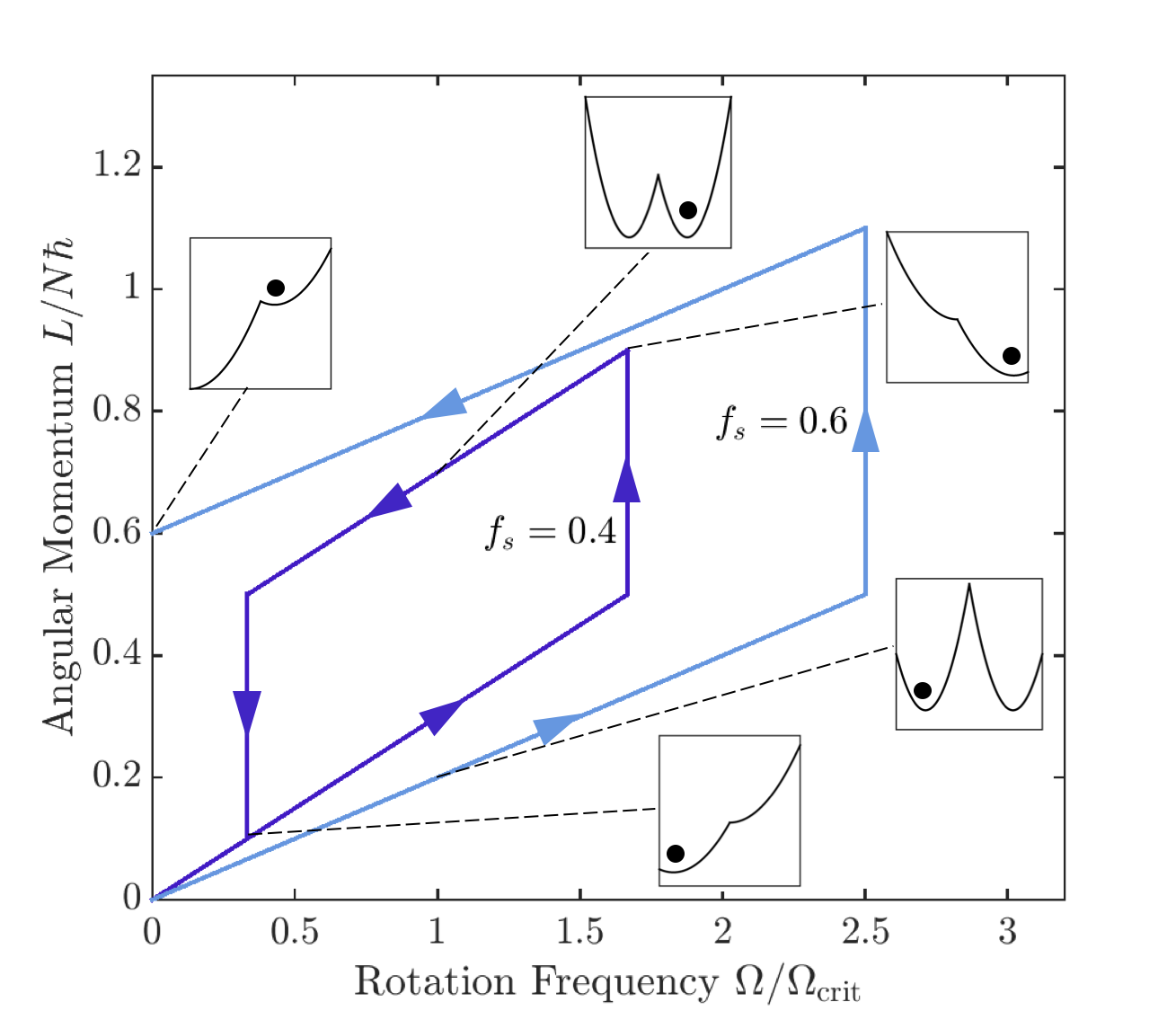}
  \caption{Schematic illustration of the change in angular momentum as the rotation frequency changes for two different values of the superfluid fraction, $f_s=0.4$ (dark blue) and $f_s=0.6$ (light blue), chosen on either side of $f_s=1/2$. The arrows indicate the increase and  decrease in $\Omega $, respectively. The insets show the ground state energy in the rotating frame as a function of angular momentum, where the bullet marks the corresponding rotational ground state.}
  \label{fig:hysteresis}
\end{figure}
It is interesting to consider the behavior of the supersolid as the rotation frequency is changed, under the conditions that energy and angular momentum are not conserved (i.e. if there was dissipation, as would be realistic in an experimental setting). Upon increasing $\Omega$ the system then first takes  angular momentum classically according to $L = I_\mathrm{solid}\Omega$. When the rotation frequency reaches $\Omega = \Omega_\mathrm{crit}$, a single vortex is energetically favorable but separated by an energy barrier, i.e., the state with just solid-body rotation is metastable against the first vortex. This metastability persists up until $\Omega = \Omega_\mathrm{crit}/(1-f_s)$, where the barrier disappears and the first vortex can enter the superfluid component. As $\Omega$ is instead decreased, we have two different trajectories depending on the value of the superfluid fraction. For $f_s < 1/2$ the condensate loses angular momentum continuously until the vortex leaves at $\Omega = \Omega_\mathrm{crit}(1-2f_s)/(1-f_s)$, i.e. there is hysteresis. On the other hand, for $f_s > 1/2$, the barrier that protects the first vortex persists. Consequently,  the condensate only loses angular momentum corresponding to solid-body rotation. This cycle of bistability is illustrated in Fig.~\ref{fig:hysteresis}. 

In conclusion, we have shown that a toroidal dipolar Bose gas may support a persistent current also in the supersolid phase. The associated minimum in the rotation energy furthermore gives rise to hysteretic behavior, similarly to what has been observed in purely superfluid ring-shaped sodium BECs~\cite{Eckel2014}.
The coexistence of a supercurrent with density modulations of the condensate establishes yet another  signature of supersolidity, that may be accessible in current experiments with dipolar gases of dysprosium or erbium. 

\bigskip
\begin{acknowledgments}
{\it Acknowledgements.} This work was financially supported by the Knut and Alice Wallenberg Foundation, the Swedish Research Council and NanoLund. Communications with G.M. Kavoulakis and P. St\"urmer are gratefully acknowledged.
\end{acknowledgments}

%\bibliography{Supersolid.bib}

\end{document}